\documentclass[seceq,amsmath,amssymb,eqsecnum]{ptptex}
\usepackage{graphicx}
\usepackage{bm}
\title{Estimation of the Lin-Yang bound of the least static energy of the Faddeev model}

\author{Minoru Hirayama$^{1}$, Hitoshi Yamakoshi$^{2}$
 and Jun Yamashita$^{1}$}

\inst{$^1$Department of Physics, University of Toyama, Gofuku 3190, Toyama 930-8555, Japan\\         
$^2$Toyama National College of Technology, Hongoumachi 13, Toyama 939-8630, Japan
}
\abst{
Lin and Yang's upper bound
$E_Q \leqq c~Q^{\frac{3}{4}}$ 
of the least static energy $E_Q$  of the Faddeev model in a sector with a fixed Hopf  index $Q$ is investigated.  By constructing an explicit trial configuration for the Faddeev field $\bm{n}$, a possible value of the coefficient $c$ is obtained numerically, which is much smaller than the value obtained quite recently by analytic discussions. }

\begin{document}

\maketitle

\section{Introduction}
It is now well known that Faddeev's $O(3)$ nonlinear $\sigma$ model (Faddeev model) possesses solitons of knot 
type \cite{L.FaddeevAJ,FaddeevAJ,Gladikowski,Hietarinta,Battye}. This model concerns the real scalar
fields 
\begin{equation}
\boldsymbol{n}(x)=(n^1(x),n^2(x),n^3(x))
\end{equation}
 satisfying
\begin{equation}
{\boldsymbol{n}}^2(x)=\boldsymbol{n}(x)\cdot\boldsymbol{n}(x)=\sum\limits_{a=1}^{3}n^a(x) n^a(x)=1. 
\end{equation}
The Lagrangian density of this model is given by
\begin{eqnarray}
{\mathcal L}_F(x) &=&c_2 l_2(x)+c_4 l_4(x),\\
l_2(x)&=&\partial_{\mu}\boldsymbol{n}(x)\cdot
 \partial^{\mu}\boldsymbol{n}(x),\\
l_4(x)&=&-\frac{1}{4}H_{\mu\nu}(x) H^{\mu\nu}(x),\\
H_{\mu\nu}(x)&=&\boldsymbol{n}(x)\cdot[\partial_{\mu}\boldsymbol{n}(x)\times\partial_{\nu}\boldsymbol{n}(x)]\nonumber\\
&=&\epsilon_{abc}n^a(x)\partial_{\mu}n^b(x)\partial_{\nu}n^c(x), \label{eqn:Hmunu} 
\end{eqnarray}
where $c_2$ and $c_4$ are constants.
Faddeev and Niemi discussed that
$\boldsymbol{n}(\boldsymbol{x})$ is intimately related to the low energy dynamics of the $SU(2)$
non-Abelian gauge field \cite{FaddeevJ.}.\\
The static energy
functional $E_F[\boldsymbol{n}]$ associated with ${\mathcal L}_F(x)$ is
given by
\begin{eqnarray}
E_F[\boldsymbol{n}]&=&\int dV [c_2 \epsilon_2(\boldsymbol{x})+c_4
\epsilon_4(\boldsymbol{x})],\\
\epsilon_2(\boldsymbol{x})&=&\sum\limits_{a=1}^{3}\sum\limits_{i=1}^{3}[\partial_in^a(\boldsymbol{x})]^2,\\
\epsilon_4(\boldsymbol{x})&=&\frac{1}{4}\sum\limits_{i,j=1}^{3}[H_{ij}(\boldsymbol{x})]^2,\\
\boldsymbol{x}&=&(x_1,x_2,x_3)=(x,y,z)
\end{eqnarray}
with $(x,y,z)$ being Cartesian coordinates. If we assume
that $\boldsymbol{n}(\boldsymbol x)$ satisfy the boundary condition
\begin{equation}
\boldsymbol{n}(\boldsymbol x)=(0,0,1) \hspace{2mm}\textrm{at} \hspace{2mm}  \arrowvert\boldsymbol{x} \arrowvert=\infty
\end{equation}
and that $\boldsymbol{n}(\boldsymbol x)$ are regular for $ |\boldsymbol{x}|<\infty$, we can regard
$\boldsymbol{n}$ as a mapping from $S^3$ to $S^2$. Such mappings are classified by the topological number $Q_H[\boldsymbol{n}]$ called the Hopf index, which is also a functional of
$\boldsymbol{n}$.  Vakulenko and Kapitanskii\cite{Vakulenko} found that the lower bound of $E_F[\boldsymbol{n}]$ is given by a constant multiple of $Q_H[\boldsymbol{n}]^{3/4}$. With the help of the best Sobolev inequality, Kundu and Rybakov found the following inequality for general $\boldsymbol{n}$ \cite{Kundu}:
\begin{eqnarray}
E_F[\boldsymbol{n}]&\geqq& K\sqrt{c_2 c_4}\biggl
 \arrowvert Q_H[\boldsymbol{n}]\biggr \arrowvert^{3/4},\\
 K&=&2^{7/2}3^{3/8}\pi^2=168.587.
\end{eqnarray}
 Then a
configuration $\boldsymbol{n}(\boldsymbol x)$ with  $Q_H[\boldsymbol{n}]\neq 0$ is stable against collapsing into a trivial configuration. 
On the other hand, Lin and Yang \cite{Lin-Yang} have shown  recently that the
least energy in a sector with a fixed $Q_H[\boldsymbol{n}]$ is bounded from the above :  
defining $E_Q$ by  
\begin{equation}
E_Q={\rm{Min}}\{ E_F[\boldsymbol{n}] \hspace{2mm}  |\hspace{2mm} Q_H[\boldsymbol{n}] = Q\},
\end{equation} 
they showed that $E_Q$  satisfies the inequality
\begin{equation}
E_Q \leqq C\sqrt{c_2c_4} Q^{3/4},
\label{eqn:LYbound}
\end{equation} 
where $C$ is a constant independent of $Q$. 
This $Q^{3/4}$ upper bound of the minimal energy is important because it ensures the stability of the configuration against collapsing into widely separated $Q$ lumps each of which has the Hopf index 1.
 As for the value of $C$, Lin and Yang did not mention. We note that Adam, S\'anches-Guill\'en, V\'azques and Wereszczy\'nski gave an analytic estimation of $C$ quite recently \cite{Adam}. 
 
For small $|Q|$, the configuration considered in ref. 5) might be a candidate which makes the energy  minimal. This configuration, however, contains the Hopf index $Q$ explicitly and the derivative of the fields$ (n_1, n_2, n_3)$ contains terms linear in $Q$. Then the energy density contains $Q^2$ and $Q^4$ terms and hence, with an appropriate choice of the scale parameter, the minimal energy becomes proportional to $Q^3$ for large $Q$. This behavior is quite different from the $Q^{3/4}$ behavior implied by the Lin-Yang theorem. In other words, it seems that the configuration other than the one considered in ref. 5)  must  be sought  to describe the minimal energy state with large $Q.$  
The purpose of  this paper is to seek a possible value of $C$.
 Since the functional $E_F[\boldsymbol{n}]$ should be  minimized for the true solution of the field equation of the Faddeev model, $E_Q$ is smaller than the value of  $E_F[\boldsymbol{n}]$ for an arbitrary trial configuration  $\boldsymbol{n}$ which has $Q_H[\boldsymbol{n}] = Q$. We calculate $E_F[\boldsymbol{n}]$ numerically for a special trial configuration whose $E_F[\boldsymbol{n}]$ is bounded above by  $ |Q_H[\boldsymbol{n}] |^{3/4}$. \\
This paper is organized as follows. After a brief explanation of the Hopf index and the static energy functional of the Faddeev model, we explain an example of the spectrum in a sector with  fixed Hopf index in Sec. II.  We thus observe that the $Q^{3/4}$ upper bound is realized only by rather special configurations.       
In Sec. III,  according to the suggestion  by Lin and Yang \cite{Lin-Yang}, we investigate a  special configuration.  It turns out that  $E_{Q}$ of this configuration indeed has the bound  $C Q^{3/4}$. We thus obtain an explicit possible value of $C$. 
The final section is devoted to  summary. We compare our numerical result with the analytic bound of $C$ given in ref. 10) and find that the value obtained in this paper is much smaller. 

%%%%%%%%%%%%%%%%%%%%%%%%%%%%%%%%%%%% 
\section{Preliminaries : Hopf index, static energy functional and example of spectrum in Hopf sector}
%%%%%%%%%%%%%%%%%%%%%%%%%%%%%%%%%%%

%%%%%%%%%%%%%%%%
\subsection{Hopf index}
%%%%%%%%%%%%%%%%
To define the Hopf index, it is convenient to introduce real fields
$\Phi_{\alpha}(\boldsymbol x)\hspace{1mm} (\alpha=1,2,3,4)$ satisfying
\begin{equation} 
\sum\limits_{\alpha=1}^{4}[\Phi_{\alpha}(\boldsymbol x)]^2=1.
\end{equation}
The complex fields $Z_1(\boldsymbol x)$, $Z_2(\boldsymbol
x)$ and a column vector $Z(\boldsymbol x)$ are defined by
\begin{eqnarray} 
Z_1(\boldsymbol x)&=&\Phi_1(\boldsymbol x)+i\Phi_2(\boldsymbol
 x),\nonumber\\
Z_2(\boldsymbol x)&=&\Phi_3(\boldsymbol x)+i\Phi_4(\boldsymbol
 x),\nonumber\\
Z(\boldsymbol x)&=&\binom{Z_1(\boldsymbol x)}{Z_2(\boldsymbol x)}.
\end{eqnarray}
If we define  the fields $n^a\hspace{1mm}(a=1,2,3)$ by
\begin{equation}
n^a(\boldsymbol x)=Z^{\dagger}({\boldsymbol x})\sigma^a Z({\boldsymbol x}),
\end{equation}
with $\sigma^a\hspace{1mm}(a=1,2,3)$ being Pauli matrices,  ${\boldsymbol n}$ is expressed as
\begin{equation}
{\boldsymbol n}=\biggl(\frac{u+u^{*}}{|u|^2+1},\frac{-i(u-u^{*})}{{|u|}^2+1},\frac{{|u|}^2-1}{{|u|}^2+1}\biggr), \label{eqn:nbyu}
\end{equation}
where the complex function $u(\boldsymbol x)$ is defined by
\begin{equation}
u(\boldsymbol x)=\biggl({\frac{Z_1(\boldsymbol x)}{Z_2(\boldsymbol x)}}\biggr)^{*}.
\end{equation}
If we define the vector potential ${\boldsymbol A}({\boldsymbol
x})=(A_1({\boldsymbol x}),A_2({\boldsymbol x}), A_3({\boldsymbol x}))$ by
\begin{equation}
A_i({\boldsymbol x})=\frac{1}{i}\{Z^{\dagger}(\boldsymbol x)[\partial_i Z(\boldsymbol x)]-[\partial_i Z^{\dagger}(\boldsymbol x)]Z(\boldsymbol x)\},
\end{equation}
we see that
$H_{ij}(\boldsymbol x)$ defined by 
\begin{equation}
H_{ij}(\boldsymbol x)=\partial_i A_j(\boldsymbol x)-\partial_j A_i(\boldsymbol x)
\end{equation}
coincides with ${\boldsymbol n}\cdot (\partial_i{\boldsymbol
n}\times\partial_j{\boldsymbol n})$ of  Eq. (\ref{eqn:Hmunu}). The Hopf index
$Q_H[{\boldsymbol n}]$ is now defined by
\begin{eqnarray}
Q_H[{\boldsymbol n}]&=&\frac{1}{16\pi^2}\int dV {\boldsymbol
 A}({\boldsymbol x})\cdot{\boldsymbol B}({\boldsymbol x}),\\
B_i(\boldsymbol x)&=&\frac{1}{2}\epsilon_{ijk}H_{jk}(\boldsymbol x).
\end{eqnarray}
Although there is no local formula expressing ${\boldsymbol
A}(\boldsymbol x)$ in terms of ${\boldsymbol n}(\boldsymbol x)$, it is known  that the Hopf index is calculated solely in terms
of ${\boldsymbol n}$. Another formula for $Q_H[{\boldsymbol n}]$ is 
\begin{equation}
Q_H[\boldsymbol n]=\frac{1}{12 \pi^2}\int dV \epsilon_{\alpha\beta\gamma\delta}\Phi_{\alpha}\frac{\partial{(\Phi_{\beta},\Phi_{\gamma},\Phi_{\delta})}}{\partial(x,y,z)},
\end{equation}
where $\epsilon_{\alpha\beta\gamma\delta}$ is the four-dimensional
Levi-Civita symbol satisfying $\epsilon_{1234}=1$. From this formula, we
can see that the allowed values of $Q_H[\boldsymbol n]$ for regular
$\Phi_\alpha(\boldsymbol x)\hspace{1mm}(\alpha=1,2,3,4)$ are integers.

%%%%%%%%%%%%%%%%
\subsection{Static energy functional}
%%%%%%%%%%%%%%%%

In terms of $u$ and $u^*$, the energy densities  $\epsilon_2(\boldsymbol{x})$ and  $\epsilon_4(\boldsymbol{x})$  are  expressed as
\begin{eqnarray}
\epsilon_2(\boldsymbol{x}) &=&\frac{4}{(1+|u|^2)^2}(\boldsymbol{\nabla} u\cdot\boldsymbol{\nabla} u^{*}),\\
\epsilon_4(\boldsymbol{x}) &=&-2\frac{({\boldsymbol \nabla}u\times{\boldsymbol \nabla}u^*)^2}{(1+|u|^2)^4}.
\end{eqnarray}
Defining $E_2[\boldsymbol{n}]$ and $E_4[\boldsymbol{n}]$ by
\begin{eqnarray}
E_2[\boldsymbol n]&=&\int dV \epsilon_2(\boldsymbol{x}),\\
E_4[\boldsymbol n]&=&\int dV \epsilon_4(\boldsymbol{x}),
\end{eqnarray}
we have 
\begin{equation}
E_F=c_2 E_2[\boldsymbol n] +c_4E_4[\boldsymbol n].
\end{equation}
From the dimension analysis, the volume integrals  $E_2[\boldsymbol{n}]$ and $E_4[\boldsymbol{n}]$ are proportional to a scale parameter $\alpha$ and its inverse $\alpha^{-1}$,  respectively. Then we have $E=\alpha \cdot c_2D_2+(1/\alpha)\cdot c_4D_4$,where $D_2$ and $D_4$ are independent of $\alpha$.
By fixing $\alpha$ appropriately, the minimum of $E$ is obtained as
\begin{equation}
E_F=\sqrt{c_2 c_4} J,\quad J=2\sqrt{D_2D_4}.
\end{equation} 
If the volume $V$ of the integral consists of some pieces $V_1, V_2, \cdots$, we have  
\begin{equation}
E_F= \sqrt{c_2 c_4}( J_1 + J_2 + \cdots), 
\end{equation}
where $J_1, J_2,  \cdots$ are obtained by taking the scale parameters $\alpha_1, \alpha_2,\cdots$ of $V_1,V_2, \cdots$ appropriately.

%%%%%%%%%%%%
\subsection{Example of spectrum in Hopf sector }
%%%%%%%%%%%%
Lin-Yang theorem concerns the minimal energy in the sector of a fixed Hopf index. To understand what type of energy spectrum is possible in such a sector, we here briefly discuss the case of the Aratyn-Ferreira-Zimerman (AFZ) configurations.\cite{AFZ}  They are the exact solutions of the model whose Lagrangian density is equal to $[l_4(x)]^{3/4}$, where $l_4(x)$ is defined in Eq.(1$\cdot $5).  The AFZ configuration is defined by
\begin{eqnarray}
u(\boldsymbol x)&=&f_{m,n}(\eta)e^{-i\psi_{m,n}},\\
f_{m,n}(\eta)&=&\frac{\cosh\eta-\sqrt{\frac{n^2}{m^2}+\sinh^2\eta}}{\sqrt{1+\frac{m^2}{n^2}\sinh^2\eta}-\cosh\eta},\\
\psi_{m,n}(\xi,\phi)&=&m\xi+n\phi,
\end{eqnarray}
where $\eta, \xi,\phi$ are toroidal co-odinates and $m$ and $n$ are integers.
For the above configuration, $Q_H[{\boldsymbol n}]$ is equal to $mn$.\\
Substituting the above $u(\boldsymbol x)$ in Eqs.(2$\cdot$11) and (2$\cdot$12) and choosing the scale parameter appearing in the definition of the toroidal co-ordinate appropriately, we find that the energy is given by
\begin{equation}
E_{m,n}=8\pi^2\sqrt{c_2c_4} \sqrt{|Q| \left(A(p)+|Q|B(p)\right)} \equiv \sqrt{c_2c_4} F(Q, p)
\end{equation}
\begin{eqnarray}
A(p)&= &\frac{(p+1)^2}{2p^2} \left[ 3p+2+(2p+1)g(p)\right] ,\\
B(p)&=& \frac{4(p+1)^3(p-1-\log p)}{p(p-1)^2},\\
g(p)&=&
\begin{cases}
\frac{\cosh^{-1}p}{\sqrt{p^2-1}} & 1< p\\
1 & p=1\\
\frac{\cos^{-1}p}{\sqrt{1-p^2}} & 0\leqq p <1.
\end{cases}
\end{eqnarray}
where $p$ is defined by
\begin{equation}
p=\left \vert \frac{n}{m} \right\vert.
\end{equation}
For fixed $Q$, the parameter $p$ can take several values.
In Fig.1, the distribution of $F(Q,p)$ is shown. 
\begin{center}
\scalebox{.7}[.7]{\includegraphics{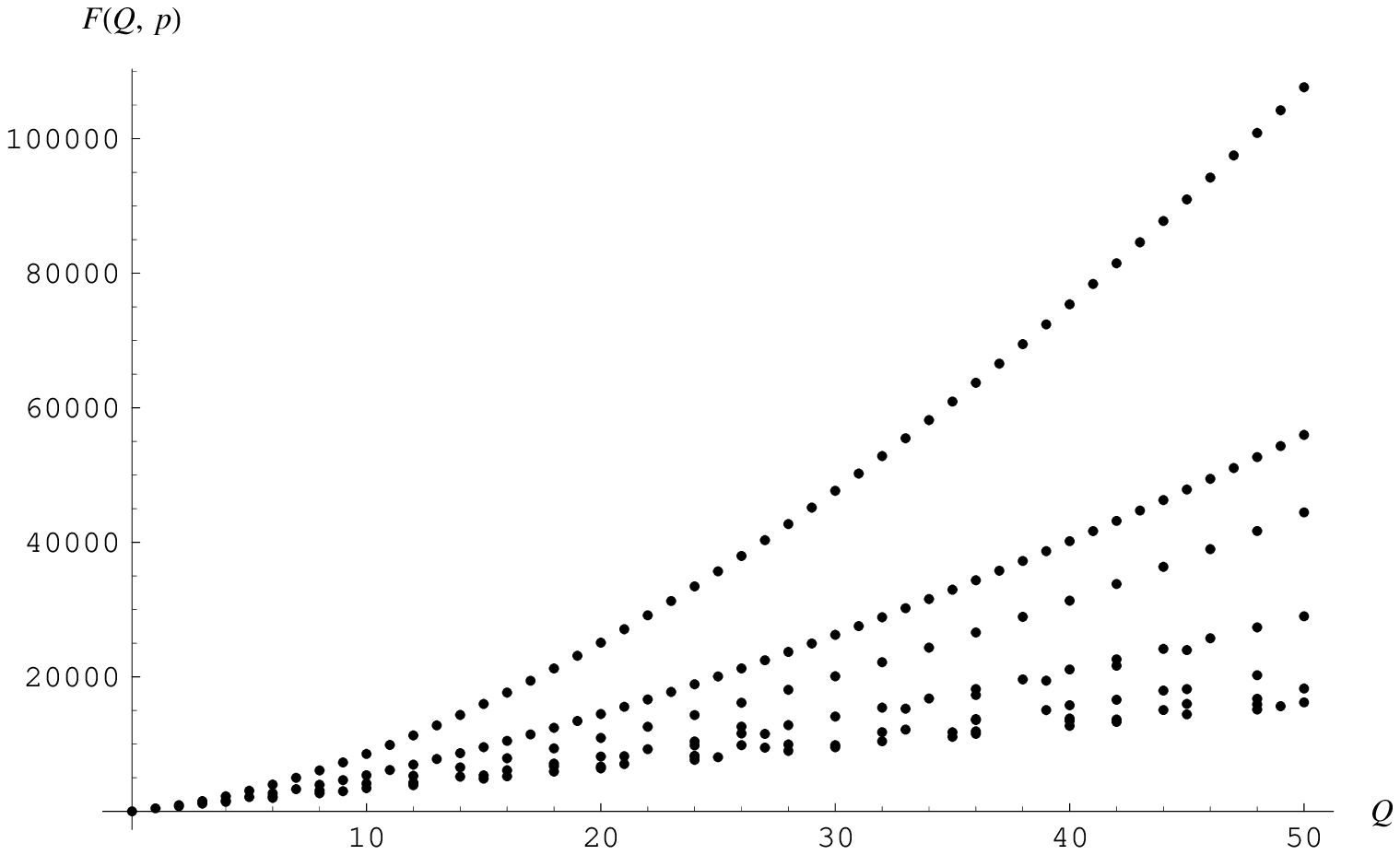}}\\
Fig.1\vspace{5mm}~~ $F(Q, p)$'s vs. $Q$.\\
\end{center}
If we denote the smallest $F(Q, p)$ in a sector with fixed $Q$ by $F_Q$, we obtain Fig. 2. \begin{center}
\scalebox{.7}[.7]{\includegraphics{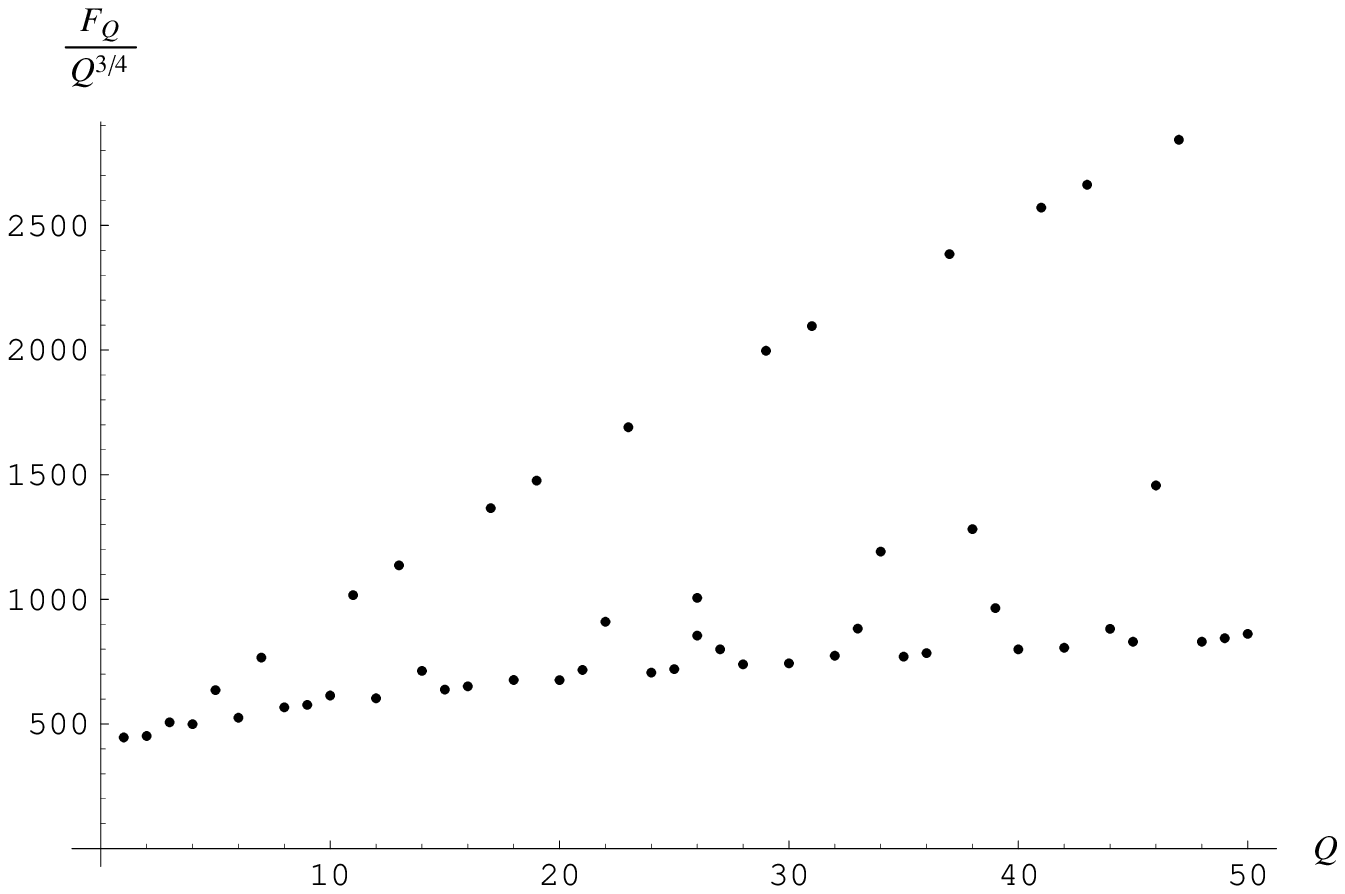}} \\
Fig.2\vspace{5mm}~~$F_Q/Q^{3/4}$ vs. $Q$.
\end{center}
Lin-Yang theorem asserts that, for the true solutions of the Faddeev model, all the points in the  $\left[F_Q/Q^{3/4}\right]$-$Q$  diagram should lie below a certain horizontal line. Off course, in the trial AFZ configurations considered here, this property is not attained.
We now proceed to consider what kind of configurations leads to the $Q^{3/4}$ upper bound.
 
%%%%%%%%%%%%%%%%%%%%
\section{$Q^{3/4}$  upper bounds}
%%%%%%%%%%%%%%%%%%%

To realize the upper bound of $E_Q$ of  the form $C Q^{3/4}$, we consider a map which is a combined map of  $g : \mathbb{R}^3\to S^3$,  $h : S^3\to S^2$ and $v : S^2 \to (S^2)'$ where $(S^2)'$ is another 2-sphere. We denote the ball in $\mathbb{R}^3 $  centered at $\boldsymbol{x}$ with the radius $\alpha$ as $B_{\alpha} (\boldsymbol{x})$. 
We assume that, for $\boldsymbol{x} \in B_\alpha(\boldsymbol{0})$, the map $g_{\alpha} : \mathbb{R}^3\to S^3$ is the stereographic projection $(x,y,z)\rightarrow(X_1,X_2,X_3, X_4)$ defined by
\begin{equation}
\left\{
\begin{array}{ll}
X_i =& {\displaystyle \frac{4f_{\alpha}(r)}{f_{\alpha}(r)^2 +4}
\frac{x_i}{r}} \ \ \quad(i=1, 2, 3), \\
X_4 =& {\displaystyle \frac{f_{\alpha}(r)^2 -4}{f_{\alpha}(r)^2 +4}}.
\end{array}
\right.
\end{equation}
Here  $f_{\alpha}(r)$ inside and outside the ball $B_\alpha(\boldsymbol{0})$ is defined by 
\begin{equation}
f_{\alpha}(r) = \left\{
\begin{array}{ll}
{\displaystyle \frac{r}{\alpha - r}} : & r < \alpha, \\
\infty : & r \geqq \alpha.
\end{array}
\right.
\end{equation}
$h$ is the Hopf map $(X_1,X_2,X_3, X_4)\rightarrow(N_1, N_2, N_3)$ defined by
\begin{equation}
u = \displaystyle{\frac{ N_1 + i  N_2}{1- N_3} = 
\frac{Z_1}{Z_2} = \frac{X_1 + i X_2}{X_3 + i X_4}}.
\end{equation}
Then $h\circ g_{\alpha} : (x,y,z)\rightarrow (N_1, N_2, N_3)$ maps $B_\alpha(\boldsymbol{0})$ to $S^2$ once and the Hopf index associated with $h$ is $1$.  We denote a point of $S^2$ by
\begin{equation}
\mathbf{N} = (\sin \Theta\cos \Phi,~\sin \Theta \sin \Phi,~\cos \Theta)
\end{equation}
 and   a point of $({\rm S}^2)'$ by 
\begin{equation}
\mathbf{N'} = (\sin \Theta'\cos \Phi',~\sin \Theta' \sin \Phi',~\cos \Theta').
\end{equation} 
We assume that  the degree of the mapping $v : S^2\to (S^2)' \quad  (\mathbf{N} \to \mathbf{N}')$ is $n$. Then the Hopf index associated with the map $v\circ h$ is equal to $n^2$ \cite{Lin-Yang}.\\ 
We define the map $w(\mathbf{r}) : \mathbb{R}^3 \to S^2$ by
\begin{equation}
w(\boldsymbol{x}) = \left\{
\begin{array}{ll}
 (v \circ h \circ g_{\alpha} )(\boldsymbol{x}) &:\boldsymbol{x} \in  B_\alpha (\boldsymbol{0}) \\
(h \circ g_{\beta})(\boldsymbol{x}- \boldsymbol{x}_i) &:\boldsymbol{x} \in B_\beta (\boldsymbol{x}_i )~(i=1, \cdots ,m) \\
(0, 0, 1) &:\rm{otherwise}.
\end{array}
\right.
\end{equation}
Here we are assuming that  $m+1$ balls are far apart from each other and do not  intersect. Then the Hopf index associated with the  map $w$ is given by \cite{Lin-Yang}
\begin{equation}
Q_H[w] = n^2+m.
\end{equation}
Considering the cases $Q_H[w]>0$, $m$ can be assumed to satisfy
\begin{equation}
0 \leqq m<2n+1
\end{equation}
 without loss of generality. 

 The map $v$ is specified by fixing $\Theta'$ and $\Phi'$ as functions of $\Theta$ and $\Phi$. It should be chosen so as to make the static energy as small as possible.
We first consider the case 
\begin{align}
\Theta' &=
\left\{
\begin{array}{ll}
0  : & 0 \leqq \Theta \leqq \frac{\pi}{4}, \\
2k\left( \Theta - \displaystyle{\frac{\pi}{4}}\right) :  & \frac{\pi}{4} < \Theta < \frac{3\pi}{4},\\
k\pi :  & \frac{3\pi}{4} \leqq \Theta \leqq \pi,
\end{array}
\right. 
\end{align}
and
\begin{equation}
\Phi' = l \Phi,
\end{equation}
where $k$ and $l$ are positive integers. For $0\leqq\Theta\leqq\pi,~0\leqq\Phi\leqq 2\pi$, $\Theta'$ and $\Phi'$ range from $0$ to $k\pi$ and from $0$ to $2l\pi$, respectively. Thus the degree of mapping of $v$ in this case is $kl$. Although it is possible to modify the above configuration so that $\Theta'$ is a smooth function of $\Theta$ even at  $\Theta = \frac{\pi}{4}\hspace{1mm}  \rm{and} \hspace{1mm}  \frac{3\pi}{4}$, we make use of the above $\Theta'$ for simplicity since we encounter no difficulty in the numerical analysis. We also note that , if the region $\frac{\pi}{4} < \Theta < \frac{3\pi}{4}$ is replaced by   $ a < \Theta <  b$ with $0<a<\frac{\pi}{4}, \frac{3\pi}{4}<b<\pi$, we would obtain a better result for the upper bound of the static energy. \\
Now, in  the case $\boldsymbol{x}\in B_\alpha(\boldsymbol{0})$,  we have
\begin{eqnarray}
\tan\Theta&= &\frac{2|Z_1| |Z_2|}{1-2|Z_2|^2}= \frac{2\sqrt{R(1-R)}}{1-2R},\\
\tan\Phi&=&\frac{X_1X_4-X_2X_3}{X_1X_3-X_2X_4}=\frac{1-S\tan\phi}{S+\tan\phi},\\
R&=&1-j \sin^2\theta,\\
S&=&r\left(\frac{1}{r-2\alpha}+\frac{1}{3r-2\alpha}\right)\cos\theta,\\
j&=&\left[\frac{4r(\alpha-r)}{r^2+4(\alpha-r)^2}\right]^2,
\end{eqnarray}
where $(r,\theta, \phi)$ are polar coordinates of $\boldsymbol{x}$. 
We also have the formulae 
\begin{eqnarray}
\epsilon'_2(\boldsymbol{x}) &= &(\nabla\Theta')^2 + \sin^2\Theta' (\nabla\Phi')^2,\\
\epsilon'_4(\boldsymbol{x})&= &\frac{1}{2}\sin^2\Theta' (\nabla\Theta'\times\nabla\Phi')^2,\\
\sin^2\Theta' &= &\frac{1}{2}\left( 1-(-1)^k\cos 4k\Theta\right),
\end{eqnarray}
\begin{equation}
\nabla\Theta'=
\left\{
\begin{array}{ll}
     2k\nabla \Theta : & \frac{\pi}{4}\leq \Theta \leq \frac{3\pi}{4}, \\
      0 : & \text{otherwise}.
\end{array}
\right.
\end{equation}
Since $\Theta$ and $\Phi$ are expressed by $R, S$ and $\phi$, we also need   
\begin{eqnarray}
J &\equiv &(\nabla R)^2\vert _{\alpha=1}=\frac{1024 r^2(1-r)^2(2-r)^2(2-3r)^2 \omega^2}{(5 r^2-8 r+4)^6},\\
K&\equiv &(\nabla S)^2\vert _{\alpha=1}=\frac{16(5 r^2-8 r+4)^2(1-\omega)}{3r^2-8r+4)^4},\\
L &\equiv &(\nabla R \cdot \nabla S)\vert _{\alpha=1}  =\frac{128 r (1-r) \omega \sqrt{1-\omega}}{(3r^2-8r+4) (5 r^2-8 r+4)^2 },
\end{eqnarray}
where $\omega$ is defined by
\begin {equation}
\omega=\sin^2\theta.
\end{equation}
Noting that $\frac{\pi}{4}<\Theta\vert _{\alpha=1}<\frac{3\pi}{4}$ corresponds to 
\begin{equation}
c_1\equiv \frac{1}{2} - \frac{1}{2\sqrt{2}}
<
\left[\frac{4r(1-r)}{5 r^2-8 r+4} \right]^2 \omega \equiv j_1(r)\omega<
\frac{1}{2} + \frac{1}{2\sqrt{2}} \equiv c_2,
\end{equation}
it is convenient to define $\int dS$ by
\begin{eqnarray} 
\int dS= \left( \int_{r_1}^{r_2} dr + \int_{r_3}^{r_4}dr\right)
     \int_{\frac{c_1}{j(r)}}^1 \frac{d\omega}{\sqrt{\omega(1-\omega)}}+
     \int_{\frac{c_1}{j(r)}}^{\frac{c_2}{j(r)}}
     \frac{d\omega}{\sqrt{\omega(1-\omega)}},
\end{eqnarray}
where $r_1,r_4$ and $r_2, r_3$ $( 0< r_1 < r_2 < r_3 < r_4< 1)$ are solutions of $j_1(r) =c_1$ and  $j_1(r) =c_2$, respectively.  Defining $X, Y$ and $Z$ by
\begin{eqnarray}
Y&=&\frac{K}{(S^2\vert_{\alpha=1}+1)^2}+\frac{1}{r^2 \omega}\nonumber\\
&=&\frac{1}{r^2\omega}+\frac{16(1-\omega)(5r^2-8r+4)^2}{\left[ (5r^2-8r+4)^2-16r^2(1-r)^2\omega \right]^2,}\\
Z&= &\frac{JK-L^2}{(S^2\vert_{\alpha=1}+1)^2}+\frac{J}{r^2\omega}\nonumber \\
&=& \frac{1024(1-r)^2(2-r)^2(3-2r)^2 \omega}{(5r^2-8r+4)^6},
\end{eqnarray}   
we obtain
\begin{eqnarray} 
\int_{B_{\alpha}(\boldsymbol{0})} dV \epsilon_2'(\boldsymbol{x})&=& \alpha\cdot [8\pi k^2 f + \pi l^2 g(k)],\\
\int_{B_{\alpha}(\boldsymbol{0})} dV \epsilon_4'(\boldsymbol{x})&= & \frac{1}{\alpha} \cdot 2\pi k^2 l^2 h(k),
\end{eqnarray}
where $f, g(k)$ and $h(k)$ are defined by
\begin{eqnarray} 
f&=&\int dS \frac{r^2\sqrt{\omega}J}{[R(1-R)]\vert_{\alpha=1}},\\
g(k)&= &\int dS\left[ 1-(-1)^k\cos 4k\Theta\right]
\left( r^2\sqrt{\omega}Y\right),\\
h(k) &= &\int dS \left[ 1-(-1)^k\cos 4k\Theta\right]
\frac{r^2\sqrt{\omega}Z}{[R(1-R)]\vert_{\alpha=1}}.
\end{eqnarray}
With an appropriate choice of the parameter $\alpha$, we find that the contribution to the static energy from the ball $B_{\alpha}(\boldsymbol{0})$ is given by 
\begin{equation} 
E_{B_{\alpha}(\boldsymbol{0})} =\sqrt{c_2c_4}D(k,l),
\end{equation}
where $D(k,l)$ is defined by
\begin{equation}
D(k,l)= 2\pi\sqrt{2} kl \sqrt{[8 k^2 f +  l^2 g(k)]  h(k)}.
\end{equation}
Similarly, the contribution to the static energy from each of the balls $B_{\beta}(\boldsymbol{x}_i)$\quad $(i=1,\cdots, m)$ is given by 
\begin{equation} 
E_{B_{\beta}(\boldsymbol{x}_i)} = \sqrt{c_2c_4} D(1,1).
\end{equation}
The total static energy is now given by
\begin{equation} 
E_{k,l} = \sqrt{c_2c_4} [D(k,l)+m  D(1,1)],
\end{equation}
while the Hopf index corresponding to the case considered here is given by 
\begin{equation} 
Q_H=(k l)^2+m.
\end{equation}
It is easy to see that  $g(k)$ and $h(k)$ have $k$-independent upper bounds. Including $f$, they are numerically calculated as
\begin{eqnarray}
f&=& 14.9,\\
g(k)&<&2\int dS r^2 \sqrt{\omega} Y=19.3\\
h(k)&<&2\int dS\frac{r^2 \sqrt{\omega} Z}{R(1-R)}=268.4.
\end{eqnarray}
We also obtain
\begin{eqnarray}
g(1)&=&9.1, \\
h(1)&=&126.5,\\
D(1,1)&=&1130.5.
\end{eqnarray}
From the upper bounds obtained above, we can discuss the upper bound of $E_Q$.  If we consider the configurations with $k=l$,  we have 
\begin{eqnarray}
E_Q &<&\sqrt{c_2c_4}\left\{2\pi\sqrt{2}k^3\sqrt{\left[8f+g(k)\right] h(k)} +D(1,1) \right\} \nonumber\\
&<&\sqrt{c_2c_4}\left[ 1711 \hspace{1mm} k^3+1131 m\right],\\
Q&=& k^4+m.
\end{eqnarray}
Here $m$ should be assumed to satisfy
\begin{equation}
0\leqq m < (k+1)^4-k^4.
\end{equation}
With the aid of the inequality
\begin{equation}
a k^3+b m \leqq (a+4b) ( k^4+m)^{\frac{3}{4}},\quad (0<a,b,k, \quad 0 \leqq m < (k+1)^4-k^4 )
\end{equation}
we find that $E_Q$ is bounded as
\begin{equation}
E_Q \leqq 6233 \sqrt{c_2 c_4} \hspace{1mm}Q^{\frac{3}{4}}.
\end{equation}
 We have thus seen that the  Lin and Yang bound is indeed realized in the configuration considered above. From this example, we conclude that the coefficient $C$ in Lin-Yang inequality should satisfy
 \begin{equation}
 C \leqq  6233 \sqrt{c_2 c_4} .
 \end{equation}

%%%%%%%%%%%%%%%
\section{Summary}
%%%%%%%%%%%%%%%
We have investigated the upper bound of the least static energy $E_Q$ of the Faddeev model in a sector with the  Hopf index $Q$. By making use of a trial configuration, we have obtained the bound
$E_Q\leqq 6233~ \sqrt{c_2c_4}~Q^{3/4}$.
Recently Adam,  S\'anches-Guill\'en,  V\'azques, and  Wereszczy \cite{Adam} discussed the same problem and gave an analytic  bound for $E_Q$ :
\begin{equation}
E_Q \leqq \frac{160~(4^3+2)~\sqrt{10} \pi^4}{3 \sqrt{2}}~\sqrt{c_2c_4}~ Q^{3/4}=7.667\cdot10^5~\sqrt{c_2c_4}~Q^{3/4}.
\end{equation} 
We see that our numerical method gives a smaller value for the coefficient of the Lin-Yang bound of $E_Q$.

\section*{Acknowledgements}
The authors are grateful to Shinji Hamamoto, Takeshi Kurimoto, Hiroshi
 Kakuhata for kind interest and N. Sawado for discussions.

\end{document}